\documentclass [PRB,twocolumn,superscriptaddress,showpacs,showkeys,preprintnumbers] {revtex4}
\usepackage[]{amssymb,amsmath,amsfonts}
\usepackage[]{graphicx}

\begin{document}

\title{Thermo-mechanical behavior of surface acoustic waves in ordered arrays of nanodisks studied by near infrared pump-probe diffraction experiments}

\author{C. Giannetti}
\affiliation{Dipartimento di Matematica e Fisica Universit\`a
Cattolica del Sacro Cuore, I-25121 Brescia, Italy}

\author{B. Revaz}
\affiliation{Dipartimento di Matematica e Fisica Universit\`a
Cattolica del Sacro Cuore, I-25121 Brescia, Italy}

\author{F. Banfi}
\affiliation{D\'epartement de Physique de la Mati\`ere Condens\'ee, Universit\`e de Gen\`eve, Gen\`eve and MANEP, Switzerland}

\author{M. Montagnese}
\affiliation{Dipartimento di Fisica,
Universit\`a degli Studi di Trieste, Italy}

\author{G. Ferrini}
\affiliation{Dipartimento di Matematica e Fisica Universit\`a
Cattolica del Sacro Cuore, I-25121 Brescia, Italy}

\author{F. Cilento}
\affiliation{Dipartimento di Matematica e Fisica Universit\`a
Cattolica del Sacro Cuore, I-25121 Brescia, Italy}

\author{S. Maccalli}
\affiliation{Dipartimento di Matematica e Fisica Universit\`a
Cattolica del Sacro Cuore, I-25121 Brescia, Italy}

\author{P. Vavassori}
\affiliation{CNR-INFM CRS S3 and Dipartimento di Fisica,
Universit\`a di Ferrara}

\author{G. Oliviero}

\author{E. Bontempi}

\author{L.E. Depero}
\affiliation{ Laboratorio di Chimica per le Tecnologie, Universit\`a degli Studi
di Brescia }

\author{V. Metlushko}
\affiliation{Department of Electrical and Computer Engineering,
University of Illinois at Chicago, Chicago, IL 60607}

\author{F. Parmigiani}
\email{fulvio@dmf.unicatt.it}
\affiliation{Dipartimento di Fisica,
Universit\`a degli Studi di Trieste\\ and Sincrotrone Trieste, I-34012 Basovizza, Trieste,
Italy }

\date{\today}
\begin{abstract}

The ultrafast thermal and mechanical dynamics of a two-dimensional lattice of metallic nano-disks has been studied by near infrared pump-probe diffraction measurements, over a temporal range spanning from 100 fs to several nanoseconds. 
The experiments demonstrate that, in these systems, a two-dimensional surface acoustic wave (2DSAW), with a wavevector given by the reciprocal periodicity of the array, can be excited by $\sim$120~fs Ti:sapphire laser pulses.
In order to clarify the interaction between the nanodisks and the substrate, numerical calculations of the elastic eigenmodes and simulations of the thermodynamics of the system are developed through finite-element analysis. 
At this light, we unambiguously show that the observed 2DSAW velocity shift originates from the mechanical interaction between the 2DSAWs and the nano-disks, while the correlated 2DSAW damping is due to the energy radiation into the substrate.

\end{abstract}
\maketitle

\section{introduction}

In recent years, the progress in the fields of transducers and sources of coherent acoustic excitations in the GHz range, triggered extensive studies on the thermo-mechanical properties of solid-state mesoscale systems. In particular, time-resolved reflectivity experiments, performed on gratings of metallic nanometric stripes (2-D confined) on transparent (SiO$_2$) or semitransparent (Si) substrates, evidenced oscillations in the GHz range \cite{Lin:1993,Bonello:2001,Antonelli:2002,Hurley:2002}, however their origin of these oscillations is still matter of debate \cite{Antonelli:2002,Hurley:2002}. The reason resides in the difficulty to model the strong mechanical coupling \cite{Lee:2005,Hurley:2006} between the nano-objects and the substrate and to calculate the mechanical eigenfrequencies of the coupled system. 

In addition, there is a lack of time-resolved experiments on the thermo-mechanical properties of 3-D confined nano-objects, mainly because of the difficulties in developing an experimental set-up with the required resolution and sensitivity. 
At the best of our knowledge, the only experiment reported in the literature concerns time-resolved reflectivity performed on arrays of gold nanometric disks on a SiO$_2$ substrate \cite{Lin:1993}, where the observed variations of the reflectivity signals were attributed to the normal modes of the single metallic nano-objects. 

In a recent letter \cite{Comin:2006}, focused on the study of the magnetic dynamics of ordered arrays of magnetic nano-disks on silicon, we suggested the possibility of triggering a collective acoustic excitation of the nano-disk array/silicon coupled
system. To demonstrate that two-dimensional surface acoustic waves (2DSAWs) can be excited in ordered arrays of metallic nanodisks, by sub-ps light pulses, the infrared diffraction time-resolved experiment reported in Ref. \onlinecite{Comin:2006} is here further extended and improved.
In the present work, by measuring the frequencies of the oscillations of the first order diffracted light, as a function of the array periodicity $\vec{D}$, we demonstrate that two-dimensional SAWs, with a wavevector $\vec{q}$ defined by $|\vec{q}|$=$2\pi/|\vec{D}|$, can be excited by a near-infrared ultrashort laser pulse. In addition, the comparison between finite-element analysis and accurate measurements of the 2DSAW velocity shift and damping, as a function of $\vec{q}$, clearly indicates that the observed 2DSAW velocity shift originates from the thermo-mechanical interaction between the 2DSAWs and the nano-disks, while the correlated 2DSAW damping is due to the energy radiation into the substrate.

The manuscript is organized as follows: Sec. \ref{set_up}, the details of the time-resolved technique are presented. In Sec \ref{results}, we report on the time-resolved diffraction measurements on samples with periodicities of 2000~nm, 1000~nm, 800~nm and 600~nm and the results are compared to the calculations of the mechanical eigenmodes. In Sec. \ref{thermodynamics} we discuss both the numerical simulations of the complete thermal evolution of the nanodisks/substrate coupled system and the model developed to account for the measured dynamics of the intensity variation of the diffracted beam. In Sec. \ref{vel_shift_damping} we address the mechanism responsible for the 2DSAW velocity shift and damping. 

\section{experimental set-up}
\label{set_up}
\label{timeresolved_spectroscopy}
\begin{figure}[t]
 \centering \includegraphics[bb=0 0 519 319, width=0.5\textwidth]{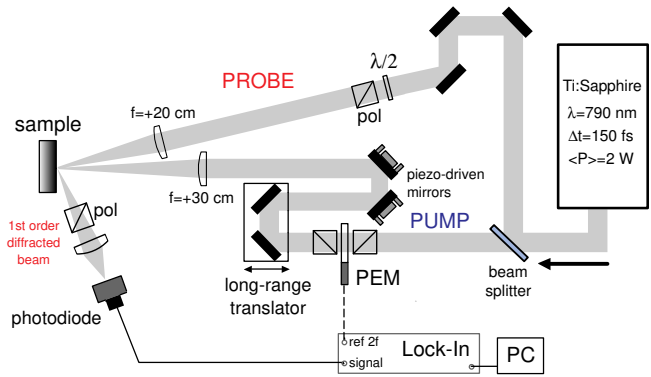}
 \caption{\small A scheme of the time-resolved optical set-up is reported. The variation of the
 intensity of the first-order diffracted probe beam is measured through a lock-in amplifier referenced at 100~kHz.}
 \label{fig:setup}
\end{figure}

The time-resolved optical set-up, sketched in Fig. \ref{fig:setup}, has been developed to achieve a high-sensitivity in very reduced perturbing conditions.

The light source is a 76~MHz repetition rate Ti:Sapphire oscillator,
operating at a central wavelength of $\simeq$790~nm and with a
pulse timewidth ($\Delta t$) of 120~fs. The 2~W laser beam is divided in the pump and
probe beams. Both the pump and probe intensities are controlled
by means of a half-wavelength ($\lambda$/2) plate and a calcite polarizer.
The energy of the pump is ~$\lesssim$10~nJ, while less than
1~nJ is used for the probe. The pump beam is modulated
at 100~kHz, through a photo-elastic modulator (PEM) placed between two
crossed polarizers. To optimize the optical signal to noise ratio, the spot sizes of the pump and probe beams was set at 60~$\mu$m and
40~$\mu$m, respectively. Second harmonic generation by a BBO crystal was used for the temporal and spatial overlap of the pump and probe beams.

The first-order diffracted probe beam, induced by the periodicity of the system, is focused and detected by a Si-photodiode. The noise due to the direct detection of the pump
beam scattered light is reduced through a crossed-polarizer, placed in front of the detector. The detected signal is
filtered by a lock-in amplifier referenced to the pump modulation
frequency. The phase of the signal is measured
by locking the internal phase of the lock-in amplifier to the
intensity modulation of the pump beam. In this way it is possible
to acquire signals in-phase and $\pi$/2-out-of-phase with respect to
the pump modulation.

\begin{figure}[t]
 \centering \includegraphics[bb=0 0 288 194,width=0.5\textwidth]{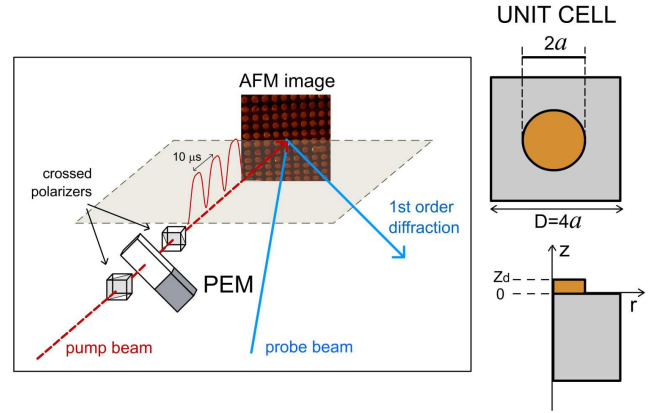}
 \caption{\small The detail of the pump beam modulation, through the PEM between two crossed polarizers, is
 shown. A modulation of the intensity of the pump beam at 100~kHz
 is obtained.
 On the right the unit cell of the nanodot array is represented.}
 \label{fig:PEMmodulation}
\end{figure}

To investigate the relaxation dynamics in the
nanosecond time-domain, a long range translator is used to change the pump
optical path up to 1.2~m, corresponding to a maximum delay
between the pump and probe pulses of $\sim$4~ns. Two spatial
points of the pump beam are imaged onto two
quadrant photodiodes interfaced to a feedback system that drives
two piezo-nano motors mounted on optical mirrors. This system is
used to keep the spatial position of the pump beam fixed during
the experiment and avoid variations of the detected signal due to
the pump-probe misalignment.

Permalloy (Py) nanosized cylinders (nanodisks), arranged in an ordered square lattice, have been synthetized by electron-beam lithography and lift-off techniques on a
Si(100) surface. The array
periods, the nanodisk thicknesses and diameters have been carefully
measured by AFM (atomic force microscopy) and grazing-angle X-ray
reflectivity. Time-resolved opical measurements have been performed on nano-disk arrays with a nominal thickness of 30~nm
and periods of 600~nm, 800~nm, 1000~nm and 2000~nm. The
diameter of the disks, as measured by AFM, is half of the array
period. A Peltier device is used to keep constant the sample back-side temperature during the experiment.

\section{Time-resolved reflectivity measurements}

\label{results}
\begin{figure}[t]
 \centering
 \includegraphics[bb=74 140 507 542,keepaspectratio,clip,width=0.48\textwidth]{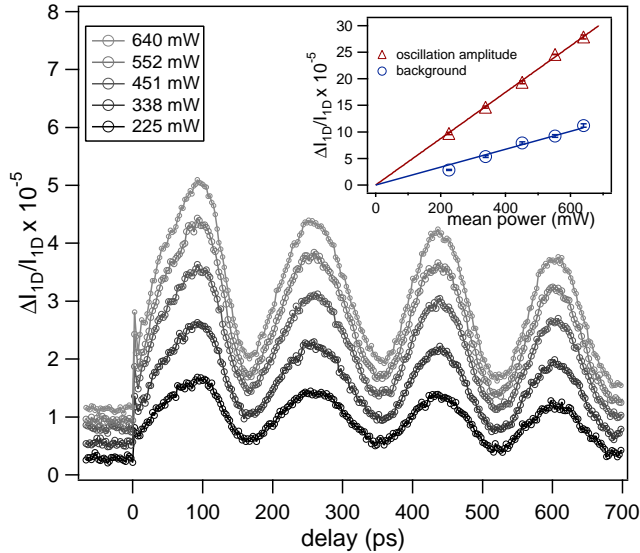}
 \caption{\small Transient reflectivity on permalloy nanodisks of 800~nm period and 400~nm diameter at different pump fluences.
 In the inset the dependence of the background level and of the oscillation amplitude is reported as a function
 of the pump intensity.}
 \label{fig:oscillazione400nm}
\end{figure}

In Fig. \ref{fig:oscillazione400nm}, the relative variation
($\Delta I_{1D}/I_{1D}$) of the intensity of the beam diffracted
from the 800~nm-period sample is reported in the sub-nanosecond
timescale. To decouple the temperature variation in the sub-nanosecond timescale from the average heating of the substrate, only the in-phase component of the variation
is displayed, see Appendix \ref{photoelastic_modulator}. $I_{1D}$ is the measured total diffracted intensity. Several sets of measurements were performed by changing the intensity of
the pump excitation. In particular, we varied the average incident
power between 225~mW (energy: 2.9~nJ/pulse) and 640~mW (energy:
8.4~nJ/pulse). 

After pump excitation at $t$=0, the signal quickly
rises and starts to oscillate with a frequency of $\sim$175~ps.
The variation of the diffracted intensity is of the order of
$\Delta I_{1D}/I_{1D}$$\simeq$1-5$\cdot$10$^{-5}$ and positive,
i.e. an increase of the intensity is detected after excitation. At
negative delays, i.e. when the probe beam investigates the sample
before excitation by the pump pulse, a positive background is
detected. This signal is attributed to the mean heating of the dot/silicon system related to the high repetition rate of the laser beam, as discussed in detail in Sec. \ref{thermodynamics}.

\begin{figure}[t]
 \centering
 \includegraphics[bb=87 120 488 755,keepaspectratio,clip,width=0.48\textwidth]{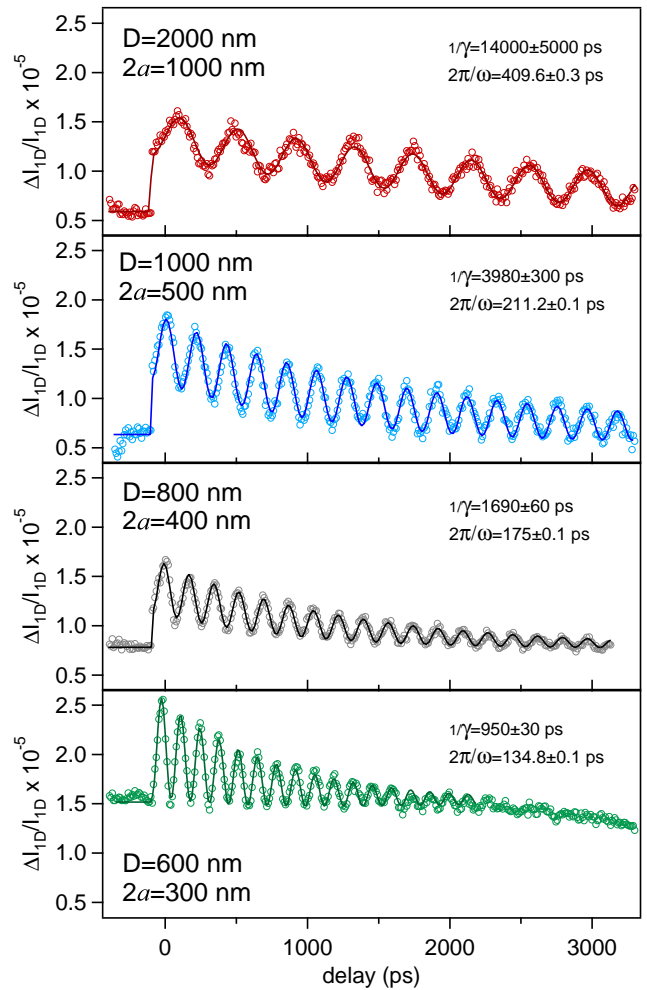}
 \caption{\small Long-range transient reflectivity (colored circles) on permalloy nanodisks arrays with different periodicity are reported. The lattice period $D$ and disk diameter 2$a$ are reported for each sample. The data has been fitted (solid lines) with Eq. \ref{result diff variation}.}
 \label{fig:q_dependence}
\end{figure}

In the inset of Fig. \ref{fig:oscillazione400nm}, the
background level (triangles) and the maximum oscillation
amplitude (circles) are reported as a function of the incident
power. The solid lines are the linear fit to data. These results
indicate that both the background and the oscillation linearly
depend on the excitation energy and, as a consequence, on the
temperature increase of the metallic array. This demonstrates that the temperature increase is small
enough to prevent non-linearities caused by the temperature
variation of the relevant optical, thermal and mechanical constants
of the system. 

\begin{figure}[t]
 \centering
 \includegraphics[bb=0 0 460 398,keepaspectratio,clip,width=0.48\textwidth]{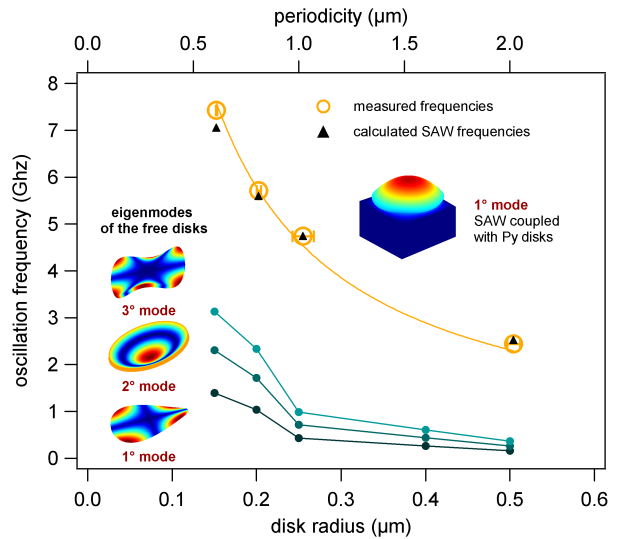}\label{SingleDisksModes}
 \caption{\small Comparison of the measured oscillation frequencies (yellow circles) to the calculated eigenfrequencies (black triangles) of the 2DSAW in silicon coupled with the metallic nanodisks, as a function of the array periodicity. The solid yellow line is the best fit of the data to a power function. The determined power exponent is $n$=-1. In addition, Fig. \ref{fig:SingleDisksModes} displays the calculated lower three eigenfrequencies (blue dots) of the isolated disks, that is decoupled from the substrate, as a function of the nanodisks radius. In all the reported calculation the measured thicknesses of the samples have been taken into account.}
 \label{fig:SingleDisksModes}
\end{figure}

To investigate the physical origin of the measured oscillation, we extended the experiment to the nanosecond timescale and measured samples with different periodicities and the same disk diameter to period ratio. From the results reported in Fig. \ref{fig:q_dependence}, we note that both the oscillation frequency and damping change as the periods are varied. In particular, we measured a variation of the
oscillation frequencies from $\sim$409~ps to $\sim$135~ps, by
changing the nominal periodicity from 2000~nm to 600~nm.
In Fig. \ref{fig:SingleDisksModes} we display the oscillation frequencies of the measurements reported in Fig. \ref{fig:q_dependence} (yellow circles), as a function of the array periodicity.
Assuming that the wavelength of the displacement is given by the periodicity of the system, the wave velocity $v_S$ is obtained performing a fit of the the data to a power function with the exponent $n$=-1. The result is $v_S=4800\pm 100$~m/s. This value is 
$\sim5\%$ below the reported SAW velocities on [001] Si surfaces
($4921$~m/s in the (100) direction and $5070$~m/s in the (110)
direction, see Ref. \cite{Pratt:1969}). Moreover, the small,
systematic shift in the measured velocity with respect to the
expected SAW velocity of a pure [001] Si surface can be explained by
the mechanical-loading due to the permalloy disks, as will be shown
below. These facts demonstrate that the oscillations in the data of
Fig. \ref{fig:q_dependence} are caused by 2DSAWs with different $q$ on
the Si surface.

To fully support this interpretation, the eigenmodes of the single
(decoupled) nanodisks have to be excluded as a possible origin for
the oscillations. To do
so, we performed a finite element analysis of the lowest three
eigenfrequencies of single nanodisks and of the 2DSAW modes, taking into account the mechanical loading
of the nanodisks. The measured sample dimensions have been used throughout the simulations.

The calculated eigenmodes are presented in
Fig.\ref{fig:SingleDisksModes}. The shape reproduces, linearly expanded, the real shape of the modes, whereas the color scale, decreasing from red to blue, refers to the modulus of the total displacement. Note that these calculations do not
require any adjustable parameters (the permalloy stiffness coefficients have been calculated from the Ni and Fe polycrystalline coefficients given in the Appendix of Ref. \onlinecite{Auld:1990}). In the case of the single
nanodisk eigenfrequencies, both the values of the frequencies and
the behavior as a function of the disks diameter differ markedly
from the experimental data. On the contrary, the simulation of the 2DSAW reproduces within few percents
the experimental frequencies. We can therefore unambiguously
conclude that the observed oscillations are caused by 2DSAW
propagating on the Si surface.

At the light of the results reported in this section, we can rationalize the physics of the system as follows. The laser-induced impulsive heating is at the origin of the disks expansion. The fast expansion of the metallic nanostructures, mechanically strongly coupled to the substrate, launches the 2DSAW in the silicon. On the subnanosecond timescale, the 2DSAW modulates the disk radius $\delta
a(t)$, inducing the measured oscillation of the diffracted signal $\Delta I_{1D}$. In the following section we develop a model to quantitatively interpret the time-dependence of $\Delta I_{1D}$.

\section{Thermodynamical and mechanical models}
\label{thermodynamics}
\subsection{Thermodynamical model}
The absorption of a sub-ps laser pulse induces an impulsive heating of the system. The larger increase of the temperature of the metallic nanodisks with respect to the substrate, triggers a spatially modulated stress on the silicon surface. The mechanical problem under investigation is therefore entangled with the thermodynamics of the system. The detailed understanding of thermo-mechanical dynamics involved in the experiments, relies on the knowledge of the spatial and temporal temperature profiles. 

A numerical simulation of the excitation and heat-diffusion process, shows that three related mechanisms must be considered. First, the average heating occurs on the milliseconds time scale and it can therefore be experimentally filtered out. Second, a nano-disk temperature increase of $\sim$ 10~K is attained on the picosecond time-scale and the related thermal expansion is responsible for launching the 2DSAW. Third, the disk-substrate heat exchange process is effective on the nanosecond time-scale. 

The numerical values of the physical constants used in this section are reported in Table \ref{TabConstants}, Appendix \ref{appendix}. Note that, in the present calculations, the boundaries of the disks are not allowed to expand, which limits the results to the thermal problem. A full thermo-mechanical analysis, which is out of the scope of the present paper, requires the inclusion in the simulations of the thermal expansion term. 

Modeling the average heating requires taking the entire sample volume into account. 
The equilibrium temperature distribution, and the timescale to attain it, are found solving the thermal problem for a
continuous laser (switch-on rise-time $\sim$100~fs, average power
$\sim$0.4~W and beam spot
radius $\sim$27.5~$\mu$m) shone on a silicon cylinder of the size of the substrate used in the actual experiment (5~mm radius$\times$ 1~mm
height; the dots volume can be neglected, being 2$\cdot$10$^{-6}$ of the total volume):
\begin{equation}\label{cont_laser}
C_{Si}(T_{Si})\frac{\partial T_{Si}}{\partial
t}=P_{con}+k_{Si}\bigtriangledown ^2T_{Si}\\
\end{equation}
where $T_{Si}$, $C_{Si}$, $k_{Si}$ are the Si lattice
temperature, heat capacity and diffusion coefficient respectively. The
power per unit volume absorbed at time $t$ and depth $|z|$ by the
Si sample from the $continuous$ laser beam is $P_{con}$(t,z,r)=$P_{con}$(t)$P_{con}$(z,r) (see appendix \ref{appendix} for the detailed analytical expression). 

\begin{figure}[t]
    \includegraphics[bb=70 70 450 300,keepaspectratio,clip,width=0.48\textwidth]{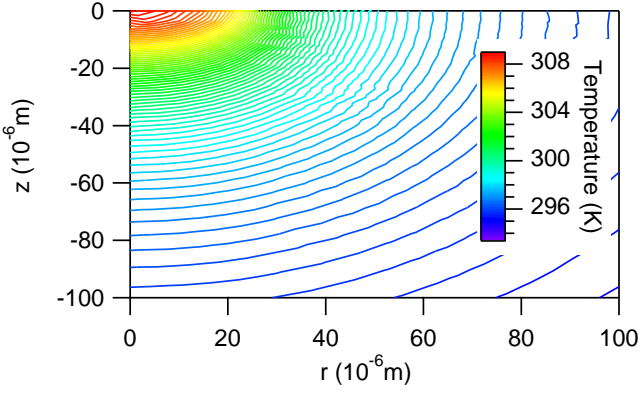}  \narrowtext
\caption{Equilibrium temperature profile of the Si slab attained 1 ms after the arrival of the first laser pulse. The temperature profile is shown from the surface down to a depth of 100 $\mu$m, vertical axis, and within a 100 $\mu$m radius from the Si substrate center, horizontal axis.}
\label{fig:Si_sub}
\end{figure}

As for the boundary conditions,
the bottom surface of the Si cylinder is set at the constant
temperature of 293.5 K, i.e. the temperature of the Peltier cell holding the sample.
The heat flux boundary condition
is imposed on all other surfaces:
\begin{equation}
\mathbf{\widehat{n}}_{Si}\cdot k_{Si}\nabla
T_{Si}=h(T_{air}-T_{Si})\label{thermal_flux_bound}
\end{equation}
$\mathbf{\widehat{n}_{Si}}$ being the unit vector normal to the substrate
surface, $T_{air}$=293.5~K the ambient air temperature, $h$ the
Si-air heat transfer coefficient. The equilibrium temperature
distribution is established after $\sim$1~ms and is shown in Fig. \ref{fig:Si_sub}. The portion of the Si slab
exposed to the laser beam has a rather constant temperature down to
a depth of 10~$\mu$m, the average temperature in this volume being
$\sim$305~K. 

The obtained result is useful for two main reasons. First, the time the period of the PEM modulation ($T$=1/100~kHz=10~$\mu$s) of the probe is at least 100 times smaller than the time necessary to heat the entire sample volume. As a consequence, we can filter out the signal related to the average heating by acquiring only the component of the diffracted signal in-phase with the pump modulation (see Appendix \ref{photoelastic_modulator}).
Second, in further determining the picosecond dynamical evolution of the electron
and lattice temperatures of the system, we will assume an initial dot and Si temperatures of 305~K.

After excitation by a single pump pulse, the thermal evolution problem spans three time scales. In the first step, the laser short pulse heats the electron gas of the metallic nanodisks (sub-picosecond time scale). In the second step, the hot electron
gas thermalizes with the lattice (picosecond timescale). In
the third step, the disks thermalize with the silicon substrate
(nanosecond time scale). This three-steps sequence repeats itself
upon arrival of a new laser pulse each 13 nsec. 

\begin{figure}[t]
    \includegraphics[bb=75 125 519 446,keepaspectratio,clip,width=0.48\textwidth]{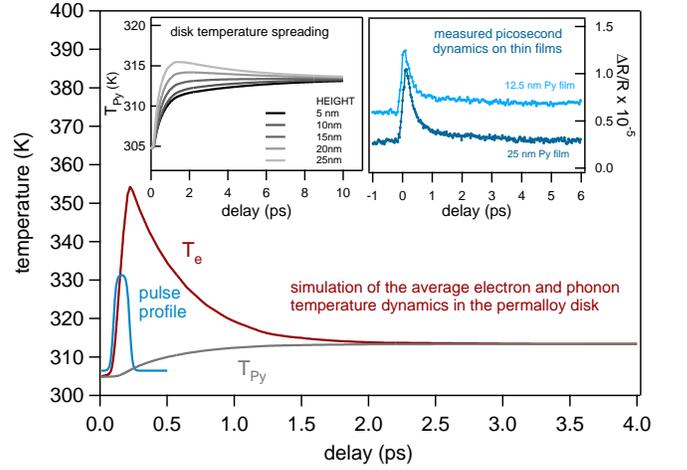}
\narrowtext
\caption{The simulation of the time evolution of the average electron temperature (T$_e$) and average lattice temperature of the permalloy disk (T$_{Py}$) is shown. The maximum
temperature of the electron system ($\sim$350~K) is reached 150~fs after laser excitation, whereas the thermalization with the lattice is completed on the picosecond timescale. An equilibrium temperature of 313~K is established $\sim$2~ps after sample excitation. The left inset displays the permalloy lattice temperature time evolution at different heights within the dot. The heights are expressed in nm with respect to the Si surface. In the right inset the time-resolved reflectivity measurements, performed on thin film samples, are reported.}
\label{fig:temperature}
\end{figure}

Finite-element analysis was performed on the unit cell reported in Fig. \ref{fig:PEMmodulation}. This simplification is justified a-posteriori, the simulations showing that there is no thermal interaction among adjacent cells, i.e. the temperature asymptotic values are established well within the cell boundaries. The unit cell boundary is actually taken as cylindrical, for sake of simplicity in writing the equations. 

The time evolution of the electron and lattice
temperatures in the nanodisks is given by the coupled equations
system \cite{Anisimov:1974}:
\begin{eqnarray}\label{2TM}\label{2TM}
C_{e}(T_{e})\frac{\partial T_{e}}{\partial
t}=-G\cdot(T_{e}-T_{Py})+P_{p}+k_{e}\bigtriangledown ^2T_{e}\\
\nonumber C_{Py}(T_{Py})\frac{\partial T_{Py}}{\partial
t}=G\cdot(T_{e}-T_{Py})+k_{Py}\bigtriangledown ^2T_{Py}
\end{eqnarray}
where $T$, $k$ and $C$ indicate temperature, thermal diffusion coefficient and heat capacity, respectively, the reference to the electron (e) or Permalloy (Py) being indicated by the subscript. $G$ is
the electron-phonon coupling constant and $C_{e}=\gamma T_{e}$. The profile of
the $pulsed$ power per unit volume absorbed by the sample is $P_{p}$(t,z,r)=$P_{p}$(t)$P_{p}$(z,r). The analytical expression for $P_{p}$(t,z,r) is reported in Appendix \ref{appendix}. 

As for the boundary conditions, the dot is thermally coupled to the underlying Si substrate via a thermal boundary resistance. The simulations show that the thermal coupling timescale is $\sim$1 ns, whereas the electron-phonon thermalization within the dot occurs on the picosecond time-scale. For this reason we discuss the sub-ps laser-electrons interaction and the ps electrons-phonons thermodynamics on the basis of equations \eqref{2TM} only, assuming the boundary with Si to be thermally insulated. The dot free boundaries are coupled to the surrounding atmosphere via convection, that is the boundary condition of the kind reported in Eq. \ref{thermal_flux_bound} applies, after the obvious substitutions for the dot electron and lattice temperatures, thermal diffusion, and unit vector defining the dot boundaries.

Simulations results are reported in Fig. \ref{fig:temperature}. The laser pulse is absorbed by the electrons, that reach a maximum temperature of $\sim$350~K in the first few hundreds of femtoseconds. The electron gas then thermalizes with the permalloy lattice $\sim$2~ps after laser excitation, the two populations attaining an asymptotic temperature $\sim$313~K. The permalloy lattice temperature evolution is calculated at different heights within the dot, the heights being expressed with respect to the Si surface. The maximum lattice temperature spread is $\sim$4.5~K and occurs during the electron-lattice thermalization process, see the left inset of Fig. \ref{fig:temperature}. This small spread is related to the finite thermal diffusion of the electrons in permalloy, smaller than in common metals.

To test the reliability of the thermodynamic simulations, we performed time-resolved reflectivity measurements on permalloy films of different thicknesses, deposited on a silicon substrate. In this case, due to the lack of spatial modulation, no SAWs are induced and the electron-phonon relaxation dynamics is more easily accessible. 
The results are reported in the right inset of Fig. \ref{fig:temperature}, for permalloy thicknesses of 25~nm and 12.5~nm. These data can be well described by Eq. \eqref{2TM}, assuming an electron-phonon coupling constant $G$ of the order of 10$^{18}$~W/(m$^3$K). This value is larger than the values measured on standard metals, such as gold and silver, but in agreement with the results previously reported for pure nickel \cite{Saito:2003}.
It is important to note that, while the reflectivity variation at negative delays and at delays larger than 2~ps is linear with the film thickness, the same signal variation is detected in the first hundreds of femtoseconds, independent of the sample thickness.
This is due to the fact that the reflectivity signal on the femtosecond time-scale depends only on the light penetration depth ($\sim$15~nm), whereas
the slow component in the transient signal, attributed to the
lattice heating, directly depends on the absorbed heat per unit volume, which is inversely
proportional to the sample thickness.  
 
At the light of these results, we can conclude that, after the excitation by the pump pulse, the temperature of the nanodisks is impulsively increased of about 8~K with respect to the equilibrium temperature. Considering that the thermal expansion of the permalloy disk induces a displacement of the silicon substrate, we can estimate that, in the first picoseconds, the displacement of the disk radius coupled to the substrate is $\delta a/a$=$\alpha \cdot \Delta T$, where $\alpha$ is the thermal expansion coefficient of the permalloy disk/silicon coupled system. Assuming $\alpha$$\simeq$$\alpha_{Si}$, being $\alpha_{Si}$=$2.6 \cdot 10^{-6}$~K$^{-1}$ \cite{Okada:1984} the Si thermal expansion coefficient at room temperature, and the temperature increase of the metallic nano-object $\Delta T\simeq$8~K, we obtain $\delta a/a$$\simeq$2.5$\cdot$10$^{-5}$.

On longer timescales, we are going to consider the metallic nanodisks as a system in thermodynamical equilibrium, which is slowly thermalizing with the cold substrate acting as a heat reservoir. In this third step the coupling between the isothermal dot and the Si substrate comes into play, via the boundary conditions:
\begin{eqnarray}\label{therm_R}
\mathbf{\widehat{n}}_{Py}\cdot k_{Py}\nabla T_{Py}+(T_{Py}-T_{Si})/R=0\\
\nonumber -\mathbf{\widehat{n}}_{Si}\cdot k_{Si}\nabla T_{Si}-(T_{Py}-T_{Si})/R=0
\end{eqnarray}
$\mathbf{\widehat{n}}_{Py}$, $\mathbf{\widehat{n}}_{Si}$ being the outward unit vector normal to the dot and Si boundary respectively and $R$ the interface thermal resistance. The interface thermal resistance is a thermodynamical parameter accounting for microscopic processes such as phonon dispersion mismatch between the disk and Si or the presence of a thin SiO$_2$ layer between the former two materials. The value for the interface resistance was taken $\sim$  10$^{-8}$ Km$^{2}$W$^{-1}$ a value compatible with values found in the literature for similar interfaces \cite{Stevens:2005,Swartz:1987}. 
         
The outcome, relevant to the present discussion, is that the thermalization occurs on the ns time scale. The detail of the coupling mechanism, however interesting, will be discussed elsewhere.

\subsection{Oscillation model}\label{sec:ex_mech}

In analogy with
the displacive excitation of coherent phonons \cite{Zeiger:1992},
we consider the equation governing the time dependence of the dot
radius, i.e. a simple harmonic oscillator equation, with a varying
equilibrium position:
\begin{equation}
\delta\ddot{a}(t)=-\omega_{0}^{2}[\delta a(t)-\delta
a_{0}(t)]-2\gamma \delta \dot{a}(t) \label{eqosc}
\end{equation}
where $\omega_{0}$ is the oscillation frequency of the surface
wave, $\delta a_{0}(t)$ is the radial variation of a
non-oscillating disk, depending on the lattice temperature and, as
a consequence, on the delay-time from the excitation pulse, and
$\gamma$ is the damping constant of the mode. We assume a direct
proportionality between the equilibrium radial variation and the
temperature increase:
\begin{equation}
\delta a_{0}(t)=l \alpha \delta T(t) \label{eqx}
\end{equation}
where $l$ is a length parameter and $\alpha$ is the thermal
expansion coefficient of permalloy. Starting from the simulations of the thermal problem described in Sec. \ref{thermodynamics}, on the nanosecond timescale we can define a unique temperature $T(t)$ of the nanodisks and assume the following
dependence:
\begin{equation}
\delta T(t)=\delta T(0) e^{-t/\tau} H(t) \label{delta}
\end{equation}
where $\delta T(0)$ is the temperature difference between the
disks and the substrate at zero delay and $\tau$ is the time
constant regulating the heat exchange between the nano-disks and
the substrate. Substituting eq. \ref{delta} in eq. \ref{eqosc}, we
obtain, in the frequency domain:
\begin{equation}
\delta a(\omega)=\frac{l \alpha \delta T_{0}
\omega_0^2}{(1/\tau+i\omega)(\omega_0^2-\omega^2+2i\gamma \omega)}
\label{eqoscw}
\end{equation}
By anti-transforming eq. \ref{eqoscw}, we obtain:
\begin{widetext}
\begin{equation}
\delta a(t)=\frac{l \alpha \delta T_{0}
\omega_0^2}{\overline{\omega}}\int_{-\infty}^{+\infty}{e^{-t/\tau}H(t)e^{-\gamma
(t-t')} \sin{\overline{\omega}(t-t')} H(t-t')dt'}
\label{soloscint}
\end{equation}
\end{widetext}
and, by solving:
\begin{widetext}
\begin{equation}
\delta a(t)=\frac{l \alpha \delta T_{0} \omega_0^2}{\omega_0^2
+1/\tau^2 -2\gamma/\tau}[e^{-t/\tau}-e^{-\gamma t}
\cos{\overline{\omega}t} + \frac{\beta}{\overline{\omega}}
e^{-\gamma t} \sin{\overline{\omega}t}]H(t) \label{solosc}
\end{equation}
\end{widetext}
where $\overline{\omega}=\sqrt{\omega_0^2-\gamma^2}$ is the
renormalized frequency and $\beta=1/\tau-\gamma$.

\begin{figure}[!]
 \centering
 \includegraphics[bb=81 92 506 548,keepaspectratio,clip,width=0.48\textwidth]{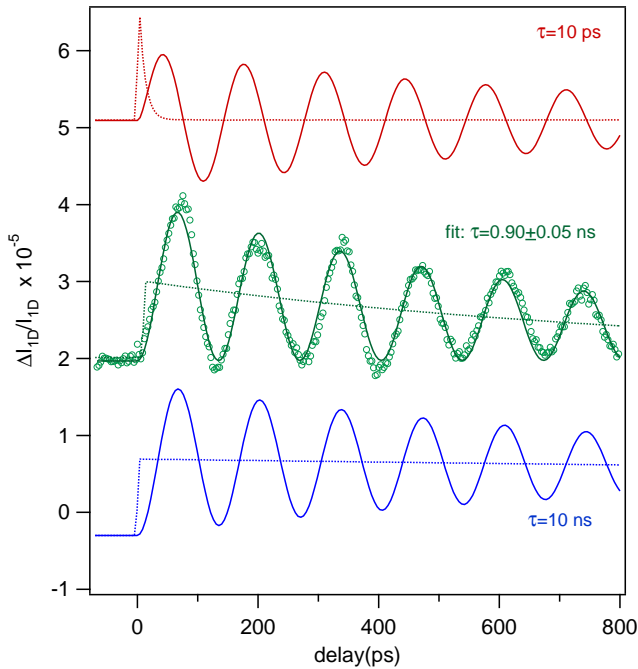}
 \caption{\small In the figure the fit (green solid line) of Eq. \ref{solosc} to the time-resolved diffracted intensity
 variation (green circles) of the 300nm-diameter sample is displayed.
 The red (top) and blue (bottom) solid lines are the expected diffracted intensity variations for impulsive and step
 relaxation dynamics, respectively. The dotted lines represents the temporal evolution of the equilibrium radius variation
 of the metallic disks.}
 \label{fig:oscillazione300nm}
\end{figure}

In Fig. \ref{fig:oscillazione300nm}, Eq. \ref{solosc} is displayed
in the limit of very short ($\tau$=10~ps, red line) and very long
($\tau$=10~ns, blue line) relaxation time constants. In the first
case, $1/\tau\sim\infty$ and $\alpha/\overline{\omega}\gg1$ and,
as a consequence, the time evolution of the oscillating mode is
dominated by the sin-like term, in contrast with the experimental
results. For this reason we can exclude, as the source of the
oscillating waves, a role played by the photo-excited electrons,
because the thermalization of the non-equilibrium electron
population acts on the femtosecond timescale. On the contrary, in
the case of very slow relaxation dynamics, the cos-like term
dominates and a damped oscillation superimposed to a step function
is expected (see Fig. \ref{fig:oscillazione300nm}). The results
reported in Fig. \ref{fig:q_dependence} indicate that the 2DSAWs are
launched by a variation of the disks radius which relaxes on a
timescale longer than the oscillation period (100-400~ps) but
comparable to the timescale probed by our set-up.

\begin{table}[t]
\caption{\label{tab:1} The oscillation periods 2$\pi/\overline{\omega}$, the inverse damping 1/$\gamma$, obtained from the fit of Eq. \eqref{result diff variation} to data, are reported. The periods of the arrays are obtained through two-dimensional FFT of the AFM images of the samples.}

\begin{ruledtabular}
\begin{tabular}{cc|c|c}
\multicolumn{2}{c}{Periods (nm)} & \multicolumn{1}{c}{2$\pi/\overline{\omega}$ (ps)} & \multicolumn{1}{c}{1/$\gamma$ (ps)}\\
\hline
\hline
$a)$&$610\pm3$   & $134.8\pm0.1$ & $950\pm30$\\

$b)$&$810\pm10$   & $175\pm0.1$ & $1690\pm60$\\

$c)$&$1020\pm50$  & $211.2\pm0.1$ & $3980\pm300$\\

$d)$&$2018\pm30$ &  $409.6\pm0.3$ & $14000\pm5000$\\
\end{tabular}
\end{ruledtabular}
\end{table}

On the basis of the previous considerations, we are now able to
evaluate the intensity variation of the first-order diffracted
beam, related to the oscillation and relaxation of the disk
radius:
\begin{equation}
\frac{\Delta I_{1D}(t)}{I_{1D}}=\frac{1}{I_{1D}}(\frac{\partial
I_{1D}}{\partial n}\delta n(t)+\frac{\partial I_{1D}}{\partial
T}\delta T(t)+\frac{\partial I_{1D}}{\partial a}\delta a(t))
\label{diff variation}
\end{equation}
where $\delta n(t)$ is the equilibrium electron distribution
variation and $\delta a(t)$ is given by Eq. \ref{solosc}. The
first term is the fast response related to the laser-induced
non-equilibrium electron population (see Sec.
\ref{thermodynamics}) and can be neglected. The second term is related to the
dependence of the optical constants of the system on the lattice
temperature and the third term takes into account the oscillation
of the dot radius. By using Eq. \eqref{diff variation q} and
\eqref{delta} in Eq. \eqref{diff variation}, we obtain:
\begin{equation}
\frac{\delta I_{1D}(t)}{I_{1D}}=K\delta T(0)e^{-t/\tau}
H(t)+\frac{2.5}{a}\delta a(t) \label{result diff variation}
\end{equation}
being $K$=$(I_{1D})^{-1}(\partial I_{1D}/\partial T$) a
coefficient related to the dependence of the permalloy
reflectivity on the temperature. Considering the impulsive expansion of the disk radius coupled to the substrate, i.e. $\delta a/a$$\simeq$2.5$\cdot$10$^{-5}$ (see Sec. \ref{thermodynamics}), we can use Eq. \ref{result diff variation} to estimate the oscillation induced in the diffracted signal by the 2DSAW excitation. The estimated oscillation amplitude $\delta I_{1D}/I_{1D}$$\simeq$6$\cdot$10$^{-5}$ is compatible with the oscillation detected in the time-resolved measurements.

By fitting the data with the
expression given by Eq. \eqref{result diff variation} it is possible
to determine two important parameters related to the mechanical properties of the nano-object array coupled with the substrate:
\begin{itemize}
\item \textbf{$\varpi$}$\simeq$$\omega_0$ is the oscillation
frequency, mainly related to the elastic constants of the
substrate \item \textbf{$\gamma$} is the damping of the 2DSAW
\end{itemize}
The data reported in Fig. \ref{fig:q_dependence} have been fitted with the function  reported in Eq. \eqref{result diff variation} and the oscillation frequency and the damping have been determined with an uncertainty drastically smaller compared to the results previously reported in the literature on similar systems \cite{Lin:1993,Hurley:2002,Hurley:2006}. In Table \ref{tab:1} we summarize the results obtained.

\section{SAW velocity shift and damping}
\label{vel_shift_damping}
2DSAWs propagate in a thin layer on the surface of the substrate, the
displacement field vanishing in the substrate within a depth of the
order of the wave length $D$. Therefore, at contrary to bulk modes,
the presence of an overlay, like the permalloy disks, is expected to
modify the properties of the surface wave. The resulting variation
of the wave velocity, in particular, is an aspect of practical
interest for SAW device applications and has been investigated by
many authors \cite{Datta:1979,Robinson:1989,Auld:1990}.
\begin{figure}[t]
 \centering
 \includegraphics[bb=0 0 442 280,clip,width=0.48\textwidth]{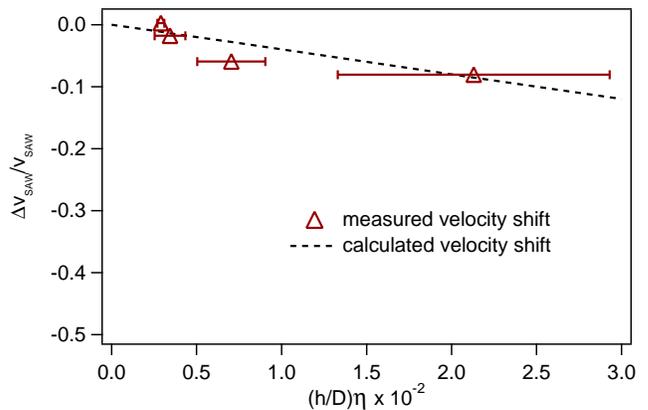}
 \caption{\small The measured 2DSAW velocity shift is reported, as a function of the parameter $(h/D)\eta$. The dotted line is the velocity shift estimated through the perturbative approach.}
 \label{fig:velocshift}
\end{figure}

We limit the present discussion to a perturbative approach presented
for example by Robinson et al. \cite{Robinson:1989}. In this
calculation, the velocity shift is proportional to the reflection
coefficient $r_s$ of the 2DSAW by the overlay, i.e. $\Delta
v_{SAW}/v_{SAW}=r_s (h/D) \eta$, where $h$ is the thickness of the
disk and $\eta$=$\pi a^2$/$D^2$ the filling factor, that is approximately $\pi/16$
for the samples of Fig. \ref{fig:q_dependence}. In the case of
permalloy on Si, it leads to $r_s=-3.81$ \cite{nota3}, to be
compared to $r_s=-1.26$ for Al on Si and $r_s=-13.6$ for Au on Si
\cite{Hurley:2002}. These values show that the density contrast
between the metallic overlay and the Si substrate is the dominant
factor determining the velocity shift (mass-loading effect). The
velocities of the 2DSAW, corrected for loading effect of the
nano-disks, are presented in the Table \ref{tab:3} and Fig.
\ref{fig:velocshift}, together with experimental values. The
calculated velocities, for which, again, no free parameter have been
used, are in reasonable agreement with the measured ones. The
variation of the experimental data at small $(h/D)\eta$ may be explained by the spread of the nanodisks thickness in the samples (see Table \ref{tab:3}).

\begin{table}[t]
\caption{\label{tab:3} Measured $v_{SAW, Meas.}$ and calculated
$v_{SAW, Calc.}$ SAW velocities. The calculated velocities have been
obtained using a perturbative approach reported in
Ref.\onlinecite{Robinson:1989}}

\begin{ruledtabular}
\begin{tabular}{cc|c|c|c|c}
\multicolumn{2}{c}{D (nm)} & 2$a$ (nm)& h (nm)& \multicolumn{1}{c}{$v_{Meas.}$ (m/s)} & \multicolumn{1}{c}{$v_{Calc.}$ (m/s)}\\
\hline \hline
$a)$&$610\pm3$& $320\pm10$& $60\pm20$ & $4525$ & $4502$ \\

$b)$&$810\pm10$& $380\pm20$&$33\pm5$  & $4629$ & $4784$\\

$c)$&$1020\pm50$&$470\pm10$& $21\pm2$ & $4834$ & $4853$\\

$d)$&$2018\pm30$&$990\pm10$&$31\pm1$ &  $4930$ & $4865$\\
\end{tabular}
\end{ruledtabular}
\end{table}

The dependence of the damping coefficient $\gamma$ is a crucial
parameter to properly address the physical mechanisms responsible
for the attenuation of the 2DSAW. In particular, a $q^4$ dependence is
expected in the case of energy radiation from the SAW to the bulk
\cite{Lin:1993}, whereas $\gamma$$\propto$$q^2$ for dissipative
damping caused by thermoelastic effect, Akhieser effect
\cite{Auld:1990} or viscosity of the electron gas \cite{Lin:1993}.

The origin of the energy radiation of 2DSAW is the following: for a
free, uniform surface, the 2DSAW is an eigenmode of the system and no
radiation is expected. When an overlay covers the surface, however,
a non-zero stress is required at the substrate-overlay interface to
force the overlay to follow the motion of the substrate surface.
This stress, more precisely its perpendicular component, is allowed
to radiate to the bulk. We follow here a discussion by Lin et al.
\cite{Lin:1993} to derive the damping expected in that case. The
perpendicular stress at the substrate-overlay interface is derived
from the equation of motion, which gives (in the limit of thin
overlays):
\begin{equation}
\sigma_{zz}=\rho_{Py}h_{Py}u_{z0}\omega^2
\end{equation}
where $\rho_{Py}$ and $h_{Py}$ are the density and height of the
permalloy film and $u_{z0}$ is the maximum displacement in the
perpendicular direction. The corresponding energy radiation can be
derived by applying a simple relation for power flow density of
acoustic waves \cite{nota4}, which leads to the following radiation
rate per unit surface $A$:
\begin{equation}
\gamma
/A\propto \frac{\rho^2_{Py}h^2_{Py}u^2_{z0}\omega^4}{\rho_{Si}v_{l,Si}}
\end{equation}
where $\rho_{Si}$ and $v_{l,Si}$ are the density and longitudinal
velocity of the Silicon substrate. To obtain a relation between the
damping and the frequency, we still have to calculate the
displacement amplitude. An estimate for this value can be obtained
by considering that, in our system, the temperature increase is kept
small enough to stay in the linear regime. We can therefore assume
that $u_{z0}$, as well, depends linearly on the temperature. Considering the same absorbed intensity, the temperature variation is inversely proportional to the mass, i.e. to the thickness of the disks. In this case, one obtains
the simple relation:
\begin{equation}
u_{z0}\propto\Delta T\propto1/h_{Py}
\end{equation}
where $\Delta T$ is the adiabatic temperature increase of the disks.
Putting together these relations leads to:
\begin{equation}
\label{damping_norm}
\frac{\pi a^2}{\gamma}\propto\lambda^4.
\end{equation}
where $\lambda$ is the 2DSAW wavelength equal to the array period.

\begin{figure}[t]
 \centering
 \includegraphics[bb=1 1 378 261,clip,width=0.48\textwidth]{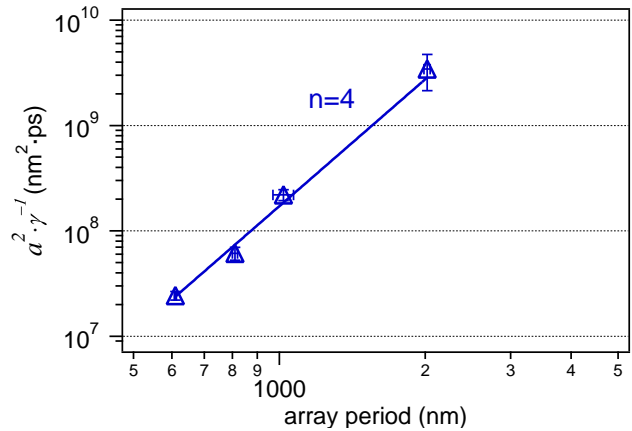}
 \caption{\small The damping normalized to the nanodisks surface is reported as a function of the array periodicity. The damping has been obtained by fitting the data reported in Fig. \ref{fig:damping} to the function given by \eqref{solosc} and \eqref{result diff variation}. The solid line is a power function with an exponent $n$=4}
 \label{fig:damping}
\end{figure}

In Fig.\ref{fig:damping} the inverse of the experimental damping rates, normalized
to the nanodisks surfaces, are reported as a function of the
frequency of the 2DSAW for the four different samples. The data can be satisfactory fitted to a power function with an exponent $n$=4, in very good agreement with the relation reported in Eq. \ref{damping_norm}.

Two conclusions can be drawn from this result: first, the radiation
of the 2DSAW to the bulk modes completely dominates the dissipative
attenuation, which may be explained by the fact that the substrate
is made of semi-conductor and crystalline material. Second, in our
thermo-mechanical system, the two assumptions used in the
calculation of the damping, i.e. the adiabaticity of the disks at
initial times and the linear relation of the displacement vs. the
temperature, result to be valid. This is an important result that
demonstrate that such a system is suitable for thermodynamics
measurements of nanoparticles.

\section{perspectives and conclusions}

The present results strongly encourage for the
development of a more detailed model of the heat exchange process between
the metallic disks and the substrate, in order to deeply investigate
the physics related to the time
constant $\tau$ regulating this process (see Section \ref{sec:ex_mech}). As discussed in this work, the mechanics and thermodynamics of the system are intimately related. The laser-induced impulsive heating of the metallic nanostructures triggers both the heat exchange with the substrate and the transfer of mechanical energy through excitation of the surface acoustic waves. The study of this physics, as the size of the nanodisks is reduced down to tens of nanometers and the temperature decreased down to a few K, could be of particular relevance in the field of the nanocalorimetry. In this regime the thermal population of the phonons in silicon is quenched and the thermal conductance can exhibit the signature of the quantization of 
the modes of the nanostructures \cite{Roukes:1999,Schwab:2000}.
It is important to underline that time-resolved experiments have already been performed on disordered metallic nanoparticles in an amorphous dielectric \cite{Nisoli:1997,Bonacina:2006}. However, in these systems, the development of a detailed model of the laser-induced impulsive heating and the following relaxation process is very difficult, as a consequence of the not well-defined thermodynamic boundary conditions at the nanoparticles/dielectric interface. 

At the same time the problem of the elastic coupling of the SAWs with the array of nanostructures deposited on the substrate is worth of investigation. In fact, the interaction of the SAWs with the periodic modulation of the surface stress can result in a deformation of the dispersion relation near the borders of the super-Brillouin zone and in the opening of a gap in the elastic surface modes. Recently it has been demonstrated that the forbidden acoustic gaps, typical of the phononic crystals \cite{Kushwaha:1993}, can be extended to the hypersonic (GHz) range in colloidal crystals formed by self-assembly of polystyrene nanoparticles \cite{Cheng:2006}. However, in this system, the modes can be excited only by thermal motion and no external control is possible. On the contrary, two-dimensional arrays of metallic nanostructures, deposited on a substrate, are the best candidates as hypersonic phononic crystals where the modes can be coherently excited and controlled by light. 

Finally, the study of the optical properties of metallic nano-objects has received
a growing attention in the last years. In
particular, when the typical size is reduced to be smaller than the
light wavelength in the visible spectrum, the
nano-objects can be represented as classical dipoles, i.e. optical
antennas, interacting with the electromagnetic field. The resonant
collective excitation of the free electron gas is responsible for
unusual and interesting electromagnetic properties such as subwavelength plasmon
resonances \cite{Barnes:2003}, super-continuum generation due to
strong field enhancement \cite{Muhlshlegel:2005,Crozier:2003},
negative permeability \cite{Grigorenko:2005} and diffraction from ordered sub-wavelength apertures in metallic surfaces \cite{Ebbesen:1998,Lalanne:2006}. A deeper understanding of
the physical processes at this length-scale is the first step
toward the development of devices in which the sub-wavelength
optics is coupled to the nano-mechanics and thermodynamics.

In conclusion, we have developed a time-resolved optical
spectroscopy in order to investigate the elastic and thermodynamical
properties of an array of metallic nano-objects deposited on a
silicon substrate. Exploiting the 2-d periodicity of the system, we
measured, directly in the time-domain, the variation of the
diffracted intensity as a function of the delay from the excitation
pulse. By changing the periodicity of the array we were able to demonstrate that surface acoustic waves are excited in
the substrate. The measurements of the 2DSAW velocity shift and damping as a function of the dimensions of the system, allowed to discuss the coupling between the 2DSAW and the nanostructures and the mechanisms responsible for the damping.
In addition, numerical calculations of the eigenmodes of the system and simulations of the thermalization process, after excitation with the pump beam, were performed through finite element analysis. Good agreement with the measurements is obtained.
These results demonstrate that metallic nano-disks arrays on semitransparent substrates are a particularly
simple model-system to study the mechanics and thermodynamics of nanometric
objects coupled to a substrate.

\appendix
\section{Average heating decoupling}
\label{photoelastic_modulator}

The pump beam is modulated through a photo-elastic modulator (PEM).
When the PEM is used between
two crossed polarizers and the maximum amplitude of the mechanical
stress corresponds to a retardation of half-wavelength of the
components along the two axis, the modulation of the light
intensity is given by:
\begin{equation}
\label{PEMmodulation} I(t)=I_0/2[1-cos(\pi sin(\omega _{P} t))]
\end{equation}
where $I_0$ is the incident light intensity and $\omega
_{P}$=2$\pi f$ is the PEM fundamental frequency. The main feature
of $I(t)$ is that the modulation is at twice the PEM frequency,
i.e. 100~kHz. 

The pump high-frequency modulation is an important
aspect for the experiment, since the detector $1/f$
noise is significantly reduced, while the system relaxation dynamics, on the picosecond
timescale, is almost unaffected by the steady-state thermal condition of the
substrate. 
This behavior is directly connected to the rate equation
governing the temperature relaxation of the system excited at
$t$$=$0 by the absorbed power per unit volume $P(t)$:
\begin{equation}
\label{DT_equation} C\frac{\partial \Delta T(t)}{\partial
t}=P(t)-k \Delta T(t)
\end{equation}
\\being $\Delta T$=$T$-$T_0$ the temperature difference from the
equilibrium temperature $T_0$, $C$ the heat capacity of the system
and $k$ the power density dissipated per unit K (neglecting the
temperature dependence of $k$). The solution of Eq.
\ref{DT_equation} is given by:
\begin{widetext}
\begin{equation}
\label{DT_equation_solution} \Delta
T(t)=\frac{1}{k}\int_{-\infty}^{+\infty}P(t-t')\frac{e^{-t'/\tau}}{\tau}H(t')dt'+\Delta
T(0)e^{-t/\tau}H(t)
\end{equation}
\end{widetext}
where $\tau$=$C/k$ is the time constant of the system, $\Delta
T(0)$ is the temperature difference at $t$=0 and $H(t)$ is the
step function.

In the limit of $\tau$$\rightarrow$0, being $\tau$ the time constant of the system, the temperature variation is given by
\begin{equation}
\label{inphase_solution} \Delta T(t)\simeq \frac{P(t)}{k}
\end{equation}
where $k$ is the dissipated power density normalized to the temperature. 

On the contrary, when $\tau \rightarrow \infty$ (average heating), the temperature variation is 
\begin{equation}
\label{outphase_solution} \Delta T(t)\simeq
\frac{1}{k\tau}\int_0^{t}P(t')dt'+\Delta T(0)H(t)
\end{equation}
being $H(t)$ the step function.
In this case, $\Delta T(t)$ is proportional to the integral of the absorbed power,
superimposed to a constant background. In particular, if the
excitation has a spectral component at a frequency $\omega$, i.e.
$P(t)$=$P_0$$cos\omega t$, Eq. \ref{outphase_solution} yields to
\begin{equation}
\label{outphase_solution2} \Delta T(t)\simeq
\frac{1}{w\tau}\frac{sin\omega t}{\omega}+\Delta T(0)H(t)
\end{equation}
Therefore, if the period of the modulation of the
absorbed power is smaller than the time necessary to the system to
dissipate the absorbed energy, the temperature modulation at the
frequency $\omega$ is damped by a factor 1/$\omega$ and
$\pi$/2-out-of-phase, with respect to the excitation.

The timescale of the relaxation dynamics of the excited system ranges up to the nanosecond, whereas the increase of the steady-state
temperature of the substrate, induced by the modulated pump intensity $P(t)$, lies in the millisecond range, as discussed in Sec. \ref{thermodynamics}.
Acquiring the 100~kHz in-phase and the
$\pi$/2-out-of-phase components of the probe signal, we decouple the temperature variation in the time-scale of interest from the average heating of the
substrate. On the contrary, standard modulation of the pump beam at 0.1-1~kHz ($T$$\sim$1-10~ms), would
result in a high in-phase background in the probe signal,
drastically decreasing the S/N ratio.

\section{Diffracted intensity variation}
\label{diffracted_intensity}

In this experiment the signal
to noise ratio can be significantly improved by measuring the
variation of the intensity of the first-order diffracted beam,
instead of the reflected beam.

The intensity of the reflected $I_{refl}$ beam from a single unit
cell of the array is given by the sum of the intensity reflected
from the dot metal surface and of the intensity reflected from the
complementary silicon surface, i.e.
\begin{equation}
\label{reflected_intensity}I_{refl}=\frac{R_{Si}(D^2-\pi
a^2)+R_{Py}\pi a^2}{D^2}I_{inc}
\end{equation}
where $R_{Si}$$\simeq$0.34 and $R_{Py}$$\simeq$0.62 are the
silicon and permalloy reflectivities at normal incidence, $D$ is
the array period, $a$ is the dot radius and $I_{inc}$ is the
intensity of the incident beam. Upon excitation with the pump
pulse, the time-dependent variation of disk radius $\delta a(t)$
induces a relative variation of the reflected signal:
\begin{equation}
\label{reflected_intensity_variation}\frac{\delta
I_{refl}}{I_{refl}}=\frac{(R_{Py}-R_{Si})2\pi
a}{R_{Si}D^2+(R_{Py}-R_{Si})\pi a^2}\delta a(t)\simeq
0.28\frac{\delta a}{a}
\end{equation}
considering that, for our samples, $D$=4$a$.

Instead, the first-order diffracted electric field is
determined by the form factor ($f_1$) of a flat disk
\cite{Hecht:1974}:
\begin{equation}
\label{f1}
f_1=\int_{S}e^{-i\mathbf{r}\cdot\mathbf{G}}ds=\int_0^{a}\int_0^{2\pi}re^{-irGcos\theta}drd\theta=\frac{2\pi}{G^2}yJ_1(y)
\end{equation}
being $S$ the dot surface, $\textbf{r}$ the position vector,
$\textbf{G}$ the reciprocal lattice vector, $r$ the radial
coordinate (parallel to the surface), $\theta$ the angle between
$\textbf{r}$ and $\textbf{G}$ and $y$=$Ga$. The intensity of the
diffracted beam is given by the sum of the intensity diffracted by
the dot array and, following Babinet's principle, of the intensity
diffracted by the negative dot array of silicon:
\begin{equation}
I_{1D}=\frac{4\pi^2}{G^4}(R_{Si}y^2 J_1(y)^2+R_{Py}y^2 J_1(y)^2)
\label{intensita diffr}
\end{equation}
With the help of the Bessel function relation:
\begin{equation}
yJ'_1(y)=yJ_0(y)-J_1(y) \label{bessel der}
\end{equation}
we obtain the relative variation of the diffracted signal:
\begin{equation}
\frac{1}{I_{1D}}\frac{\partial
I_{1D}}{\partial a}\delta a=2G\frac{J_0(y)}{J_1(y)}(R_{Si}+R_{Py})\delta
a\simeq 2.5\frac{\delta a}{a} \label{diff variation q}
\end{equation}

In the diffraction configuration, the S/N ration is thus increased
by a factor 2.5/0.28$\simeq$9. Due to the fact that the relative
intensity variation of the diffracted beam is of the order of
10$^{-5}$, the increase of a factor 9 in the S/N ratio is
necessary to obtain reliable measurements.

\section{Space-time profile of the absorbed power}
\label{appendix}
We report the detailed expressions for the power absorbed per unit volume, the boundary conditions and the  numerical values for the parameters used in the simulations for the thermal problem addressed in section \ref{thermodynamics}.

The space-time profile for the power per unit volume absorbed by the Si sample from the $continous$ laser used in eq. \ref{cont_laser} is
\begin{eqnarray} \nonumber P_{con}(t)= 1-\exp(t/\tau)\\
P_{con}(z,r)=I_{0,con}\frac{(1-R_{Si})}{\lambda_{Si}}\exp(z/\lambda_{Si})H(r_{0}-r)
\end{eqnarray}
where $I_{0,con}$= 1.68$\times$10$^{8}$ Wm$^{-2}$ is the equivalent
average $continuous$ laser power per unit surface, $\tau$=100 fs
the laser switch-on rise-time, $\lambda_{Si}$=6.37$\times$10$^{-6}$
and $R_{Si}$=0.332 the penetration depth and normal incidence
reflection coefficient for Si at a wavelength of 800 nm
respectively, calculated using the values of $n_{Si}$ and $k_{Si}$
reported in Tab. \ref{TabConstants}. A hard-edge laser beam spot size of radius r$_{0}$ = 27.5 $\mu$m is assumed, this is formally accounted for by the Heavyside function $H(r-r_{0})$. 
The z-coordinate is zero on the Si surface and
becomes negative within the Si bulk (see the inset Fig. \ref{fig:PEMmodulation}).

The space-time profile for the $pulsed$ power per unit volume absorbed by the sample, used in eq. \ref{2TM}, is
\begin{widetext}
\begin{eqnarray}
\nonumber P_{p}(t)= \frac{1}{\sqrt{2\pi}}\exp[(t-t_{0})^2/2\tau_{p}^2]\\
P_{p}(z,r)=I_{0,p}\frac{(1-R_{Si})}{\lambda_{Si}}\exp(z/\lambda_{Si})\times
[H(r-a)-H(r-r_{0})]+\\
\nonumber I_{0,p}\frac{(1-R_{Py})}{\lambda_{Py}}\exp[(z-z_{d})/\lambda_{Py}]\times
H(a-r)\times[H(z)-H(z-z_{d})]
\end{eqnarray}
\end{widetext}
$I_{0,p}$=3.83$\times$10$^{13}$ W/m$^2$ being the laser intensity, and $\tau_{p}$=120 fs the pulse width. The penetration depth and reflection coefficient at normal incidence for the Permalloy at a wavelength of 800 nm are $\lambda_{Py}$=17$\times$10$^{-9}$
and $R_{Py}$=0.617 respectively, both calculated using the values of $n_{Py}$ and $k_{Py}$
reported in Tab. \ref{TabConstants}. The first term in $P_{p}$(t,z,r) accounts for laser absorption by the disk, the second sum accounts for absorption by the uncovered portion of the Si substrate. Laser absorption by the Si substrate portion covered by the disks is neglected, being the disk's thickness $a\geq \lambda_{Py}$.

\begin{table} [t]
\caption{Constants used in solving the thermal problem addressed in section in section \ref{thermodynamics}.}\label{TabConstants}
\begin{ruledtabular}
\begin{tabular}{c|cc|c}
$\gamma_{Py}$           & 1.065$\times$10$^{3}$ & (Jm$^{-3}$K$^{2}$)   & \cite{Allen:1987}  \\
C$_{e}$                 & 3.248$\times$10$^{5}$ &(Jm$^{-3}$K$^{-1}$)          & $\gamma_{Py}$$\times$305 K  \\
C$_{Py}$                & 2.2$\times$10$^{6}$   &(Jm$^{-3}$K$^{-1}$)   				& \cite{Beaurepaire:1996}  \\
C$_{Si}$                & 1.5$\times$10$^{6}$   &(Jm$^{-3}$K$^{-1}$)   & \cite{Okhotin:1972}  \\
k$_{e}$                 & 60                   &(Wm$^{-1}$K$^{-1}$)   & \cite{nota2}  \\
k$_{Py}$                & 13-20                  &(Wm$^{-1}$K$^{-1}$)   & \cite{nota1}  \\
k$_{Si}$                & 130                    &(Wm$^{-1}$K$^{-1}$)   & \cite{Glassbrenner:1964}  \\
G                       & 8$\times$10$^{17}$    &(Wm$^{-3}$K$^{-1}$)   & \cite{Anisimov:1974} \\
R$_{th}$                & 10$^{-8}$             &(W$^{-1}$m$^{2}$K)    & \cite{Stevens:2005},\cite{Swartz:1987} \\
\textit{n}$_{Py}$       & 2.25                  &                      & \cite{Beaurepaire:1996} \\
\textit{k}$_{Py}$       & 3.6                   &                      & \cite{Beaurepaire:1996} \\
\textit{n}$_{Si}$       & 3.72                  &                      & \cite{Beaurepaire:1996} \\
\textit{k}$_{Si}$       & 0.01                  &                      & \cite{Beaurepaire:1996} \\
$h$                       &5.6                    &(Wm$^{-2}$K$^{-1}$)   & \cite{Stocker:1999}\\
\end{tabular}
\end{ruledtabular}
\end{table}
\bibliography{bibliodots}

\begin{thebibliography}{42}
\bibitem[{not({\natexlab{d}})}]{nota1}
\bibinfo{note}{The value for k$_{e}$ has been estimated on the basis of the
  Wiedemann-Franz equation \cite{Pastori:2000} k$_e$=$\sigma$$L$T, where
  $\sigma$=7.69$\times$10$^{6}$ $\Omega^{-1}m^{-1}$ is the electrical
  conductivity for the Permalloy \cite{Landolt:1986}, $L$=2.45$\times$10$^{-8}$
  W$\Omega$k$^{-2}$ the Lorenz constant and T the temperature in Kelvin.}
\end{thebibliography}

\end{document}